%% file: main.tex
\documentclass{article}
\usepackage{spconf,amsmath,graphicx}
\makeatletter
\renewcommand{\section}{%
  \@startsection{section}{1}{\z@}%
  {6pt}
  {3pt}
  {\bf\centering\uppercase}
}
\renewcommand{\subsection}{%
  \@startsection{subsection}{2}{\z@}%
  {4pt}%
  {2pt}%
  {\bf}%
}
\renewcommand{\subsubsection}{%
  \@startsection{subsubsection}{3}{\z@}%
  {3pt}%
  {1pt}%
  {\it}%

}
\renewcommand{\thebibliography}[1]{%
  \section{References}%
  \list{[\arabic{enumi}]}{%
    \settowidth\labelwidth{[#1]}%
    \leftmargin\labelwidth
    \advance\leftmargin\labelsep
    \itemsep 2pt
    \parsep 0pt
    \parskip 0pt
    \usecounter{enumi}%
  }%
  \def\newblock{\hskip .11em plus .33em minus .07em}%
  \sloppy\clubpenalty4000\widowpenalty4000%
  \sfcode`\.=1000\relax%
}
\makeatother
\usepackage{booktabs} 
\usepackage{tabularx}
\usepackage{graphicx}
\usepackage{subcaption} 
\usepackage{caption}
\usepackage{cite}
\usepackage{comment}
\usepackage{amsmath,amssymb,amsfonts}
\usepackage{algorithmic}
\usepackage{graphicx}
\usepackage{textcomp}
\usepackage{xcolor}

\title{Deformable 2D Gaussian Splatting for efficient 4k Video Compression}
\name{Chenhao Zhang,  Fengqing Zhu}
\address{Purdue University, West Lafayette, Indiana, U.S.A.
}
\begin{document}
%
\maketitle
\begin{abstract}
Ultra-High-Definition (UHD) video presents significant challenges for efficient storage and real-time decoding. Learning-based methods, such as Neural Video Compression (NVC) and Implicit Neural Representations (INR), achieve competitive rate–distortion performance but suffer from high decoding latency and excessive memory usage. Meanwhile, Gaussian Splatting has recently attracted attention in the computer graphics community due to its ultra-fast rendering and high-fidelity visual quality. Despite these advantages, its application in video compression remains largely unexplored. To bridge this gap, we propose a real-time video compression framework that represents and compresses a Group of Pictures (GOP) using a coarse-to-fine multi-scale 2D Gaussian Splatting (2DGS) structure coupled with a lightweight deformation network. Experiments demonstrate that our method delivers rate-distortion performance in LPIPS that surpasses H.265 and other state-of-the-art learning-based video compression methods. Our work demonstrates the potential of Gaussian Splatting as a practical solution for efficient high-resolution video compression.
\end{abstract}
\input{sections/introduction}
\input{sections/related_work}

\input{sections/method}

\input{sections/experiment}

\input{sections/conclusion}

\vspace{-3mm}
{\fontsize{9.8pt}{11.8pt}\selectfont
\bibliographystyle{IEEEbib}
\bibliography{strings,refs}
}

\end{document}

%% file: sections/introduction.tex
\section{Introduction}
\label{sec:intro}

The demand for Ultra-High-Definition (UHD) video is growing rapidly, driven by popular streaming services, cloud gaming, and remote collaboration. As 4K video tends to become the standard, storing and transmitting this massive amount of data has become a major challenge. While traditional video codecs\cite{AVC,HEVC,VVC} have effectively exploited data redundancy through hand-crafted architectures, their performance gains are gradually saturating. The increasing complexity required to achieve marginal compression improvements indicates diminishing returns under the traditional coding paradigm.

To break this ceiling, the community has pivoted toward learning-based approaches. Neural Video Compression (NVC) methods\cite{DVC,DCVC} leverage large-scale datasets to capture video characteristics in a data-driven manner, while Implicit Neural Representations (INR) \cite{NeRV,hinerv} offer a novel paradigm by mapping spatio-temporal coordinates directly to RGB values. Although these learning-based methods have demonstrated Rate-Distortion performance comparable to or even surpassing state-of-the-art traditional codecs, they suffer
\begin{figure}[t]
    \centering
    \includegraphics[width=\columnwidth]{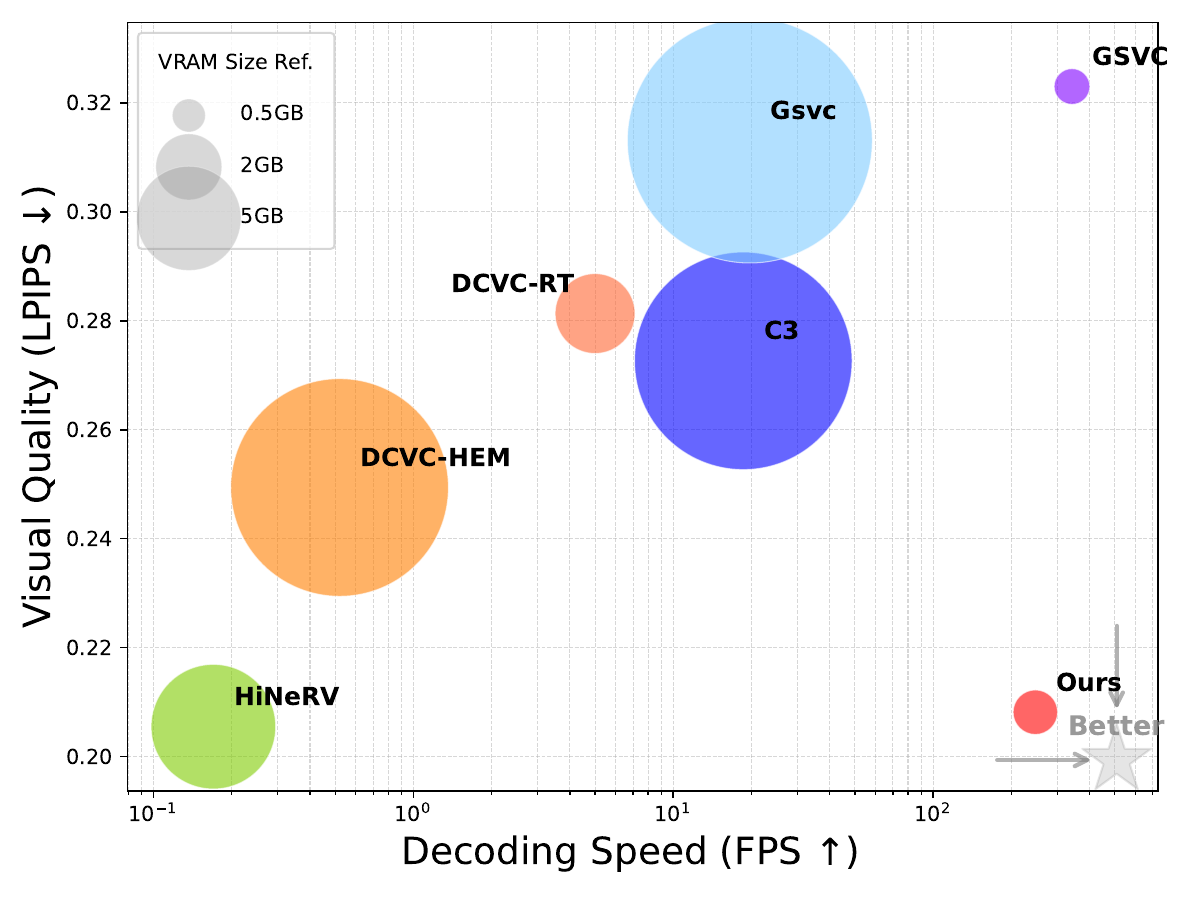}
    \vspace{-2em} 
    \caption{Perceptual quality (LPIPS) vs. decoding speed (FPS) at a uniform bitrate of 0.1. The bubble size represents VRAM consumption. Our method achieves the best trade-off between visual fidelity and decoding efficiency.}
    \vspace{-2em} 
\end{figure}
from severe practical impediments. Specifically, NVC frameworks typically incur prohibitive computational costs, leading to substantial decoding latency and high VRAM usage. These factors create a major bottleneck for their adoption on consumer-grade platforms. While INRs alleviate the memory footprint to some extent, they still fall short of real-time decoding due to the heavy inference overhead of MLPs. Adding to these challenges, computational demands increase rapidly with image dimensions, which poses a formidable barrier for UHD video applications where efficiency is critical.

In parallel, the computer graphics community has witnessed significant progress with the emergence of Gaussian Splatting (GS) \cite{3DGS}. Although initially designed as a rendering method, GS functions as a signal representation similar to INRs. Crucially, empowered by highly optimized rasterization technique, it offers ultra-fast rendering speeds and efficient memory usage—precisely bridging the aforementioned gap of decoding efficiency in high-definition video compression. Despite these compelling advantages, the application of GS in the specific context of video compression remains in a nascent stage. 

To address this research gap and overcome the inherent limitations of existing NVC and INR based video compression methods, we propose a real-time and highly efficient video compression framework built upon multi-scale 2D Gaussian Splatting (2DGS), which is particularly well-suited for high-resolution video content.
Our approach utilizes a coarse-to-fine 2DGS coupled with a multi-plane deformation network to achieve an extremely compact representation for a Group of Pictures (GOP). We leverage optical flow guidance for the deformation network and design a multi-stage training scheme to effectively reconstruct the temporal information within the GOP. In addition, quantization-aware training (QAT) is adopted to maintain the stability and robustness of the model under quantized inference.

Our contributions are as follows:
\begin{itemize}
    \item We introduce a novel multi-scale 2DGS architecture integrated with multi-plane deformation, enabling an highly efficient representation of UHD video content.
    \item We employ an optical flow-guided multi-step training strategy to guarantee the effective and accurate representation of motion and dynamics within the GOP.
    \item Our framework enables efficient, real-time decoding for 4K videos, achieving up to 247 FPS and less than 1GB VRAM usage. Moreover, it delivers rate–distortion performance comparable to H.265 and competitive with state-of-the-art deep learning–based video compression methods.
\end{itemize}

\begin{figure*}[ht!]
    \centering
    \vspace{-0.3cm}
    \includegraphics[width=0.95\textwidth]{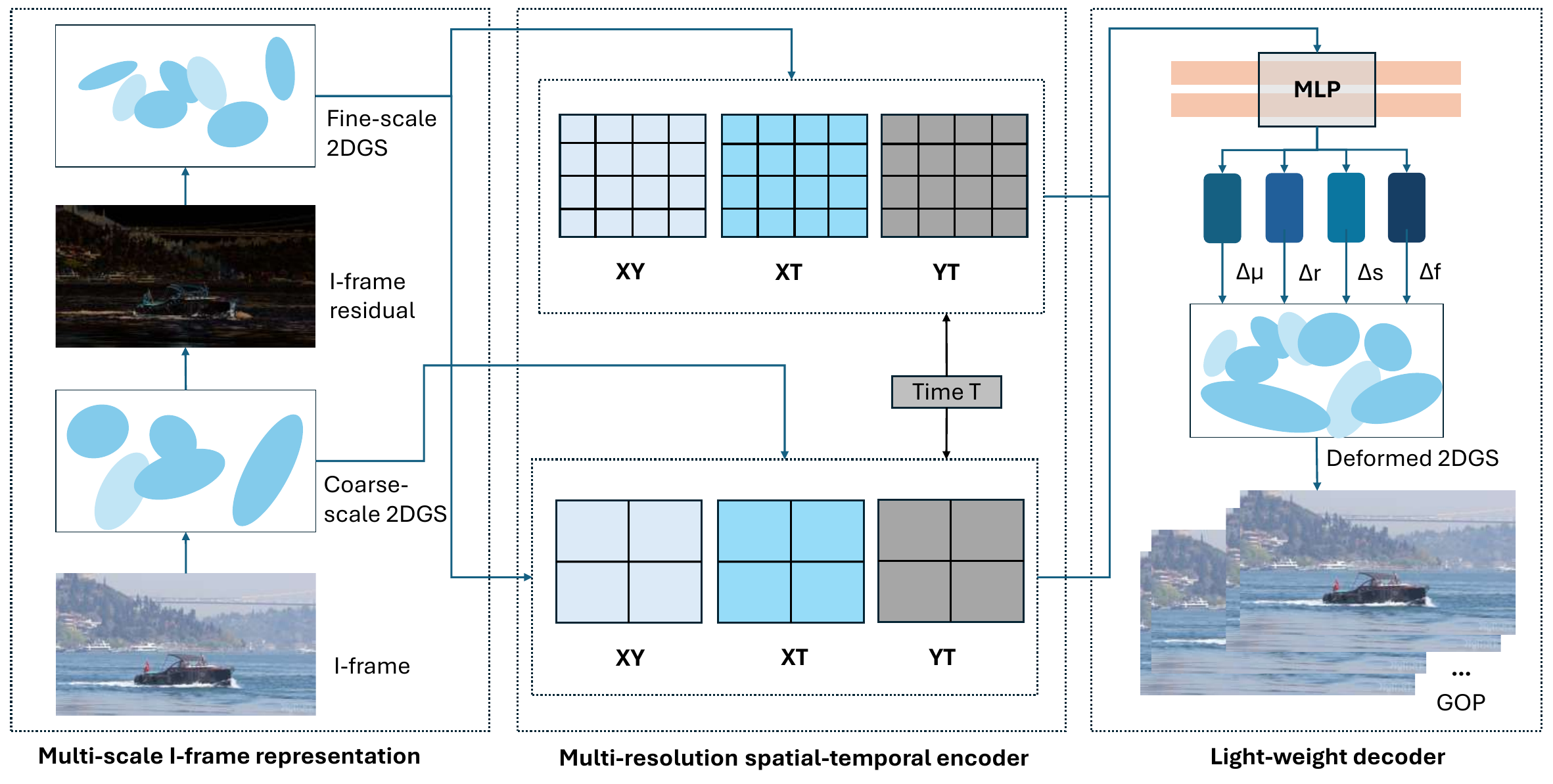}
    \vspace{-0.2cm}
    \caption{The overall framework of the proposed method: A multi-scale 2DGS representation (left) is first established for I-frame. Subsequently, a deformation network comprising a plane-based encoder (middle) and a lightweight decoder (right) is employed to warp the 2DGS primitives to subsequent frames. }
    \label{fig: overall}
    \vspace{-0.4cm}
\end{figure*}

%% file: sections/related_work.tex
\section{related work}
\label{sec:rel_work}

\subsection{Neural Video Compression}
Traditional video codecs rely on complex, hand-crafted architectures to  eliminate redundancy. With the rise of deep learning, some methods began using neural networks to replace specific components, such as in-loop filtering \cite{8736997} and pre/post-processing \cite{sandwich_my}. End-to-end paradigms, such as DVC~\cite{DVC} and the DCVC series~\cite{DCVC,DCVC_HEM}, take a significant step further by jointly optimizing the entire compression pipeline. Driven by the rate-distortion loss~\cite{balle2017proceedings}, these methods achieve superior performance that surpasses traditional codecs like H.265 and H.266. While end-to-end NVC models offer significant gains, their heavy reliance on complex neural network for feature extraction and motion estimation results in substantial memory overhead and latency~\cite{10.24963/ijcai.2025/1165}. Even with the specialized efficiency designs in DCVC-RT\cite{DCVC_RTimage}, there remains a considerable gap before seamless UHD video compression can be realized on consumer-level edge platforms. 

\subsection{INR-based Video Representation}
Unlike Neural Video Compression (NVC) frameworks that rely on explicit feature extraction and motion compensation, Implicit Neural Representations (INR) parameterize continuous signals within coordinate-based neural networks, a concept pioneered by NeRV \cite{NeRV}. This approach shifts the focus from traditional video coding to a neural model compression challenge. As a result, mature optimization techniques—such as weight pruning, quantization, and entropy coding~\cite{han2016deepcompressioncompressingdeep}—can be seamlessly integrated to achieve compact representations.

Although state-of-the-art INR frameworks~\cite{hinerv,hnerv} have achieved rate-distortion performance on par with some standardized codecs under certain configurations, they face inherent bottlenecks such us inference latency, particularly at Ultra-High-Definition (UHD) resolutions, ultimately hindering their adoption in real-time scenarios.

\subsection{GS-Based Video Representation and Compression}
Gaussian Splatting (GS)~\cite{3DGS} has emerged as a disruptive rendering technique using Gaussian primitives for high-fidelity, low-latency synthesis. GaussianImage~\cite{zhang2024gaussianimage} simplified this for image compression, while D2GV~\cite{D2GV,11148004} introduced deformation fields for video representation. In the compression domain, GSVC~\cite{10.1145/3712678.3721876} represents frames individually with 2DGS, ensuring low latency but neglecting temporal correlations. Conversely, Gsvc~\cite{GSVC_NIPS} utilizes 3DGS for entire scenes, but adapting multi-view 3D primitives to 2D video introduces significant spatial redundancy. Consequently, applying GS for efficient UHD video compression remains an under-explored frontier.

%% file: sections/method.tex
\section{method}
\label{sec:method}

\subsection{Preliminary}
\noindent\textbf{Gaussian Splatting} 3D Gaussian Splatting (3DGS) \cite{3DGS} represents a 3D scene into a group of oriented Gaussian ellipsoids. Each Gaussian ellipsoid is characterized by the 3D center point $\boldsymbol{\mu}\in \mathbb{R}^{3}$ and a covariance matrix $\mathbf{\Sigma}\in \mathbb{R}^{3 \times 3}$.

The color of the Gaussian ellipsoid is parameterized by a spherical harmonics (SH) coefficients vector $c$ and opacity $\alpha$.  Given a fixed camera view, 3DGS renders by approximating the projection of all the Gaussian ellipsoids along the depth dimension into the image pixel coordinate. 
The final pixel color $\mathbf{C} \in \mathbb{R}^3$ is determined by $\alpha-blending$ $N$ Gaussians on the image plane:
\begin{equation}
    \mathbf{C} = \sum_{i=1}^{N} c_i \alpha_i \prod_{j=1}^{i-1} (1 - \alpha_j ), 
    \label{eq:volume_rendering}
\end{equation}
where $c_i$ is the color contribution of the $i$-th Gaussian. The weight $\alpha_i $ incorporates the opacity $o_i \in (0, 1)$ and the spatial influence of the $i$-th Gaussian.

GaussianImage\cite{zhang2024gaussianimage} simplifies this process by reducing each attributes to 2D, with a covariance matrix $\mathbf{\Sigma} \in \mathbb{R}^{2 \times 2}$ and a center point $\boldsymbol{\mu} \in \mathbb{R}^2$ to get 2DGS. Image-GS\cite{zhang2025image} furtherly reduce the reliance on opacity by directly accumulating each 2DGS's color contributions during rendering and maintain a angle $\theta$ for the rotation.  Finally, 2D Gaussian primitive is fully characterized by 4 trainable parameters:
\begin{equation}
    \mathbf{G} = (\boldsymbol{\mu}, \mathbf{r}, \mathbf{s}, \mathbf{f})
\end{equation}
which makes it a highly compact representation well suitable for compression-oriented applications.
\subsection{Overview}
A straightforward but naive way to represent a video with 2DGS is to independently fit a separate set of Gaussians for each frame, as done in early work such as GSVC\cite{10.1145/3712678.3721876}. This approach, however, ignores temporal correlations and introduces substantial redundancy. Inspired by traditional video codecs—where P-frames reference I-frames within a Group of Pictures (GOP)—we propose to encode only a single reference frame per GOP using 2DGS, and to reconstruct all remaining frames by warping this set of Gaussians. Following deformation-field–based representations in computer graphics \cite{D3DGS}, we treat the reference frame’s 2DGS parameters as a canonical space and use a neural network to model their temporal evolution. To further enhance reconstruction quality for UHD content, we introduce a multi-scale 2DGS architecture trained in a coarse-to-fine manner and incorporate optical flow as a motion prior when optimizing the coarse layer.
\subsection{Multi-scale Representation}
\noindent\textbf{Multi-scale 2DGS} For a given Group of Pictures (GOP) containing $N$ frames, denoted as $\mathcal{F} = \{f_1, f_2, \dots, f_N\}$, we designate the middle frame $f_{N//2}$ as our Intra-frame ($f_I$). This keyframe $f_I$ is represented using a muiti-scale 2D Gaussian Splatting architecture. We assume a fixed total number of Gaussians $M$ and then We utilize a predefined weight coefficient $\lambda$ ($\lambda \in (0, 1)$) to partition $M$ into coarse and fine sets $G_c, G_f$. We first use $\lambda M$ Gaussians to overfit $f_I$ by minimizing the photonic loss between the rendering result of coarse scale 2DGS $render(G_c)$ and $f_I$. Due to the limited count, these Gaussians are only able to mainly capture the low-frequency components of $f_I$. 
The reconstruction of the low-frequency image $f'_I$ is thus defined as:
\begin{equation}
    f'_I = \text{render}(G_c)
\end{equation}
We then utilize the remaining $(1 - \lambda)M$ Gaussians to overfit the residual between the original keyframe and its low-frequency reconstruction $|f_I-f_I'|$. This forces the fine scale  Gaussians $G_f$ to capture the high-frequency details that the coarse layer struggled to represent. This coarse-to-fine multi-scale architecture enables us to effectively leverage 2DGS for the highly efficient representation of high-resolution images.

\noindent\textbf{Multi-scale Deformation} 
We then utilize a deformation network to warp the canonical 2DGS sets $G_c$ and $G_f$ to all other frames in the GOP. We adopt the Multi-Plane\cite{kPlanes} structure as our encoder to efficiently capture both temporal and spatial details. We have one spatial plane $XY \in \mathbb{R}^{H\times W\times C}$ and two spatial-temporal planes $XT \in \mathbb{R}^{H\times D\times C}$, $YT \in \mathbb{R}^{W\times D\times C}$. Where  $W$, $H$, and $D$ denote the feature resolutions for the spatial width, height, and temporal depth respectively and $C$ represents the feature dimensions. We then take Hadamard product of these features across planes to aggregate them back to 3D:
\begin{equation}
    \mathbf{F}(\mathbf{q}) = \mathbf{F}_{XY} (q_x, q_y) \circ \mathbf{F}_{XT} (q_x, q_t) \circ \mathbf{F}_{YT} (q_y, q_t)
\end{equation}
Where $q_x$,$q_y$ and $q_t$ is the querying 2DGS coordinates.
To adapt this structure to our Multi-Scale 2DGS architecture, we employ a novel hierarchy of multi-resolution planes to achieve scale-specific feature encoding. Specifically, we construct two distinct sets of Multi-Planes operating at different grid resolutions: a low-resolution plane set $\mathcal{P}_{L}$ and a high-resolution plane set $\mathcal{P}_{H}$. To accommodate the hierarchical nature of our representation, we employ a scale-aware feature assignment strategy. For the coarse 2DGS set $\mathcal{G}_c$, which maintains the foundational motion structure, we exclusively query the low-resolution planes $\mathcal{P}_{L}$. In contrast, for the fine 2DGS set $\mathcal{G}_f$, designed to capture high-frequency details, we adopt a progressive fusion approach by concatenating features from both the low-resolution $\mathcal{P}_{L}$ and high-resolution $\mathcal{P}_{H}$ planes. Let $\mathbf{F}_{L}(\mathbf{q})$ and $\mathbf{F}_{H}(\mathbf{q})$ denote the queried features at coordinate $\mathbf{q}$. 

Subsequently, the assembled feature vector $\mathbf{h}$ is fed into a unified, extremely lightweight MLP decoder $\Psi$ to interpret the temporal dynamics. This decoder predicts the frame-wise deformation for the Gaussian attributes:
\begin{equation}
    \Delta \boldsymbol{\mu}, \Delta r, \Delta \mathbf{s}, \Delta f = \Psi(\mathbf{h}(\mathbf{q}))
\end{equation}
Here, $\Delta \boldsymbol{\mu} \in \mathbb{R}^2$, $\Delta r \in \mathbb{R}$, $\Delta \mathbf{s} \in \mathbb{R}^2$, and $\Delta f \in \mathbb{R}^3$ represent the deformations applied to the 2D position, rotation angle, scaling, and color coefficients, respectively.
\subsection{Optical-flow Guided Multi-stage Training}
Leveraging optical flow always facilitates the effective capture of temporal dynamics. In order to maintain compatibility with our multi-scale representation, we propose a staged training paradigm. 

In the first stage, optical flow is incorporated as an auxiliary supervision signal to train the coarse-layer 2DGS and the low-resolution grid encoder. Since the coarse layer is designed to represent low-frequency global content, this supervision enables the model to rapidly establish a robust global motion trajectory. The objective function in this stage integrates the photometric loss $\mathcal{L}_{rgb}$, an optical flow alignment loss $\mathcal{L}_{flow}$, and a grid smoothness regularization $\mathcal{L}_{smooth}$:
\begin{equation}
    \mathcal{L}_{stage1} = \mathcal{L}_{rgb} + \lambda_{f}\mathcal{L}_{flow} + \lambda_{s}\mathcal{L}_{smooth}.
\end{equation}
The photometric loss $\mathcal{L}_{rgb}$ is defined as a weighted sum of the $\mathcal{L}_1$ loss and the D-SSIM loss:
\begin{equation}
    \mathcal{L}_{rgb} = (1-\lambda)\mathcal{L}_1 + \lambda \mathcal{L}_{D-SSIM}.
\end{equation}

Upon establishing the fundamental motion framework, we proceed to the second stage where explicit optical flow supervision is detached to avoid the propagation of estimation noise. We then perform joint optimization of the fine-layer 2DGS and the high-resolution grid encoder. The training focus shifts toward capturing high-frequency spatial-temporal details, governed by the loss function:
\begin{equation}
    \mathcal{L}_{stage2} = \mathcal{L}_{rgb} + \lambda_{s}\mathcal{L}_{smooth}.
\end{equation}
\noindent\textbf{Quantization Aware Training} To achieve further model size reduction while ensuring quality preservation, we implement quantization aware training to simulate the effects of quantization. Specifically, for each quantized attribute, uniform noise sampled from the range $[-\frac{\Delta}{2}, \frac{\Delta}{2}]$ is introduced, where $\Delta$ is defined as the quantization step size. This process effectively models the discretization error, making the entire model more robust to quantization. Since the quantization operation (e.g., rounding) is non-differentiable, we use straight-through estimator (STE)\cite{bengio2013estimating} to facilitate gradient flow during training. And we use arithmetic coding\cite{arith} to turn all the quantized paramters into bitstreams.

%% file: sections/experiment.tex
\begin{figure*}[ht!]
    \centering
    \vspace{-0.3cm}
    \includegraphics[width=0.95\textwidth]{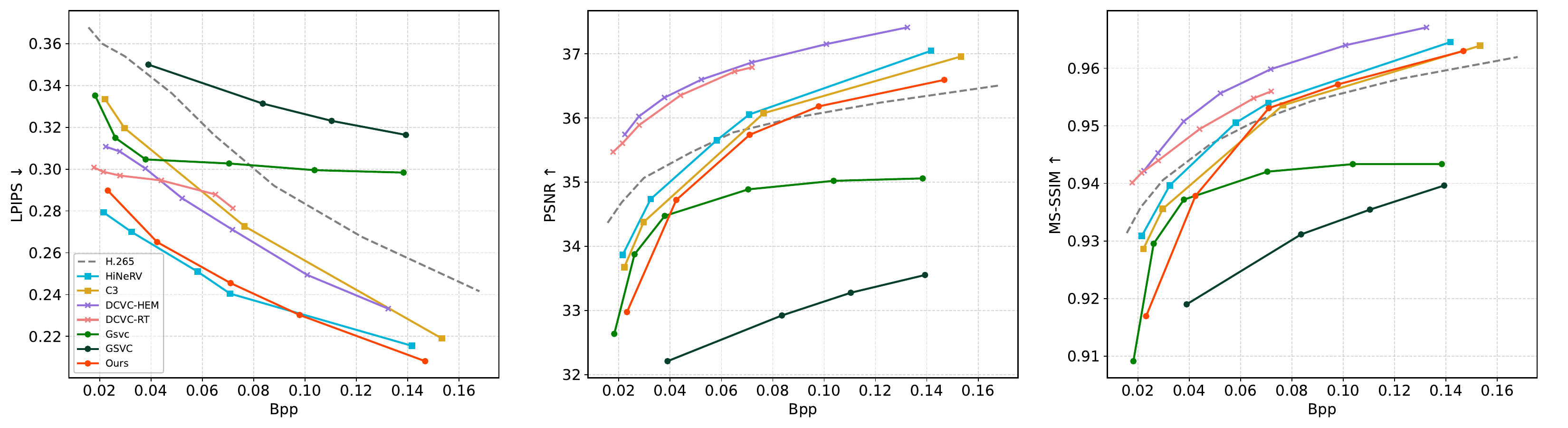}
    \vspace{-0.2cm}
    \caption{Rate-distortion curves on UVG\cite{UVG} dataset}
    \label{fig: RD}
    \vspace{-0.4cm}
\end{figure*}
\section{Experiment}
\subsection{Experimental setting}
\noindent\textbf{Dataset} We conduct our experiments on the widely used UVG dataset (4K, YUV 4:2:0, 8-bit, 120FPS, first 120 frames) \cite{UVG}. We perform all the experiments (training and testing) in RGB space.

\begin{table}[ht!]
\centering
\footnotesize
\setlength{\tabcolsep}{4pt}
\caption{Compression efficiency comparison on UVG\cite{UVG}.}
\label{tab:performance_comparison}
\begin{tabular}{l c c c c} 
\toprule
\textbf{Method} & \textbf{Bpp} & \textbf{Enc FPS $\uparrow$} & \textbf{Dec FPS $\uparrow$} & \textbf{VRAM (MB) $\downarrow$} \\ 
\midrule
DCVC-HEM\cite{DCVC_HEM} & 0.038 & 0.31    & 0.53  & 21832 \\
HiNeRV\cite{hinerv}   & 0.037 & 0.001     & 0.17  & 5073    \\
C3 \cite{C3}      & 0.039     & 0.00002     & 18.58     & 20909
    \\
DCVC-RT\cite{DCVC_RTimage}  & 0.038     & 9.41     & 10.13     & 2941    \\
GSVC \cite{10.1145/3712678.3721876}    & 0.043     & 0.0006    & 322.64    & 464    \\
Gsvc \cite{GSVC_NIPS}    & 0.038    & 0.0009    & 21.46     &  14276   \\
Ours     & 0.041 & 0.005     & 227.39 & 824 \\
\midrule
DCVC-HEM\cite{DCVC_HEM} & 0.101     & 0.31     & 0.53     & 21832    \\
HiNeRV\cite{hinerv}   &  0.097    & 0.0009     & 0.15    & 7154   \\
C3 \cite{C3}      & 0.102     & 0.00002     & 18.59     & 21810    \\
DCVC-RT\cite{DCVC_RTimage}  & -     & -     & -     & -    \\
GSVC \cite{10.1145/3712678.3721876}     & 0.105     & 0.0006     & 260.39     &  488  \\
Gsvc \cite{GSVC_NIPS}    & 0.103     & 0.0009     & 19.74    & 27646    \\
Ours     & 0.103     & 0.004     & 202.     & 912   \\
\bottomrule
\end{tabular}
\end{table}

\noindent\textbf{Baselines} We compare our method with the state-of-the-art INR-based method: HiNeRV \cite{hinerv},C3\cite{C3}, the end-to-end trained neural video codec: DCVC-HEM \cite{DCVC_HEM}, DCVC-RT\cite{DCVC_RTimage}, conventional codec H.265 via FFmpeg, and two recent GS-based methods: Gsvc \cite{GSVC_NIPS} and GSVC \cite{10.1145/3712678.3721876}.

\noindent\textbf{Metrics} For evaluating video reconstruction quality, we report two widely used distortion-based metircs: PSNR and MS-SSIM \cite{MSSSIM} and one perceptual metric: LPIPS\cite{lpips}. For evaluating compression efficiency, we measure the encoding/decoding frames per second (FPS) and peak Video Random Access Memory (VRAM) consumption.
\subsection{Implement Details} We implement our framework upon the open-source GaussianImage\cite{zhang2024gaussianimage} repository. All experiments are conducted on a single NVIDIA RTX 5000 Ada GPU. We set the GOP size to be 10. We employ Videoflow\cite{videoflow} to generate the optical flow for supervision in first stage training.

\begin{figure}[h!t]
    \centering
    \begin{subfigure}{0.48\columnwidth}
        \centering
        \includegraphics[width=\linewidth]{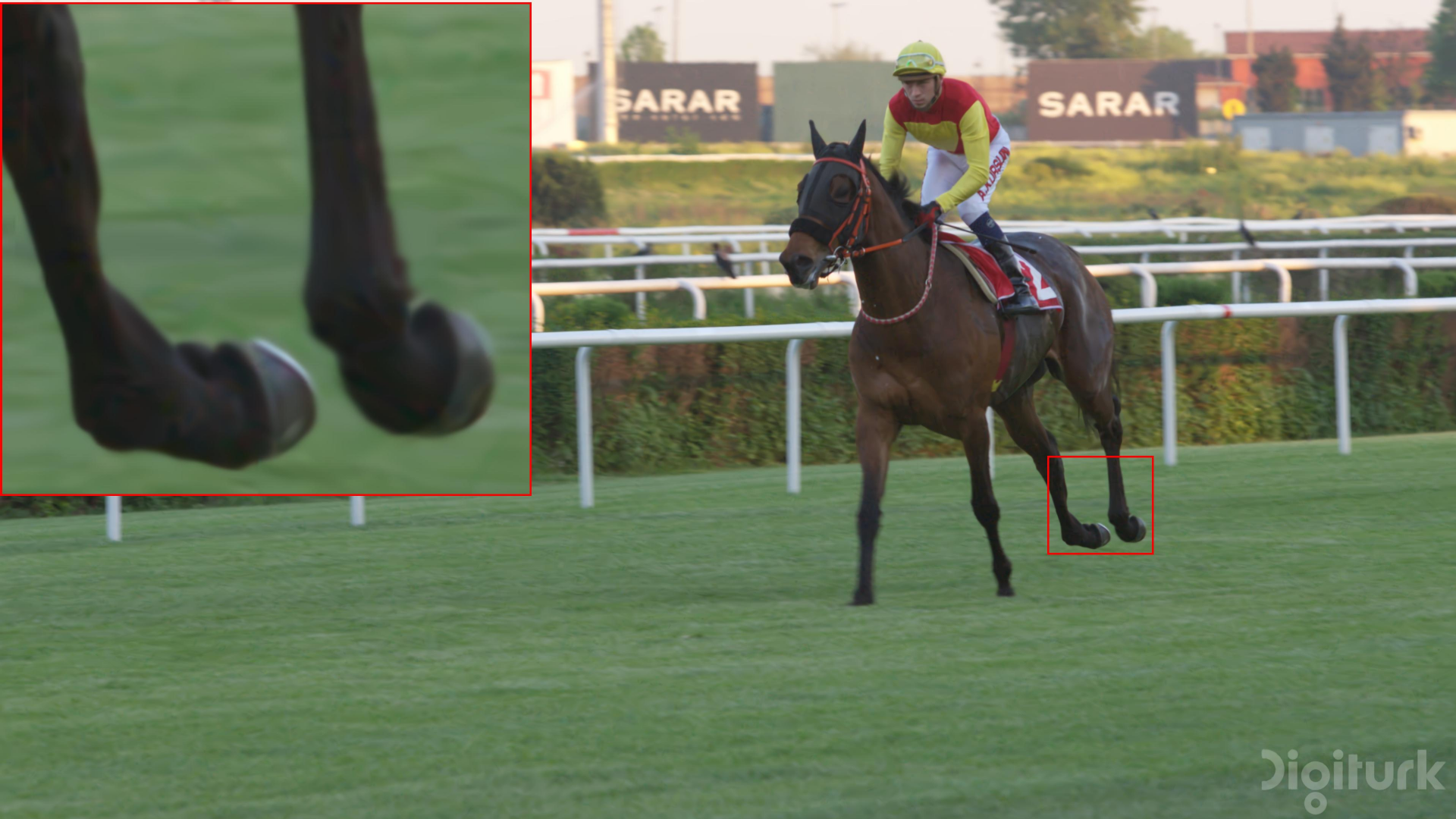}
        \caption{GT\\ \phantom{Bpp: 0.000, LPIPS: 0.00}   }
        \label{fig:result1}
    \end{subfigure}
    \hfill
    \begin{subfigure}{0.48\columnwidth}
        \centering
        \includegraphics[width=\linewidth]{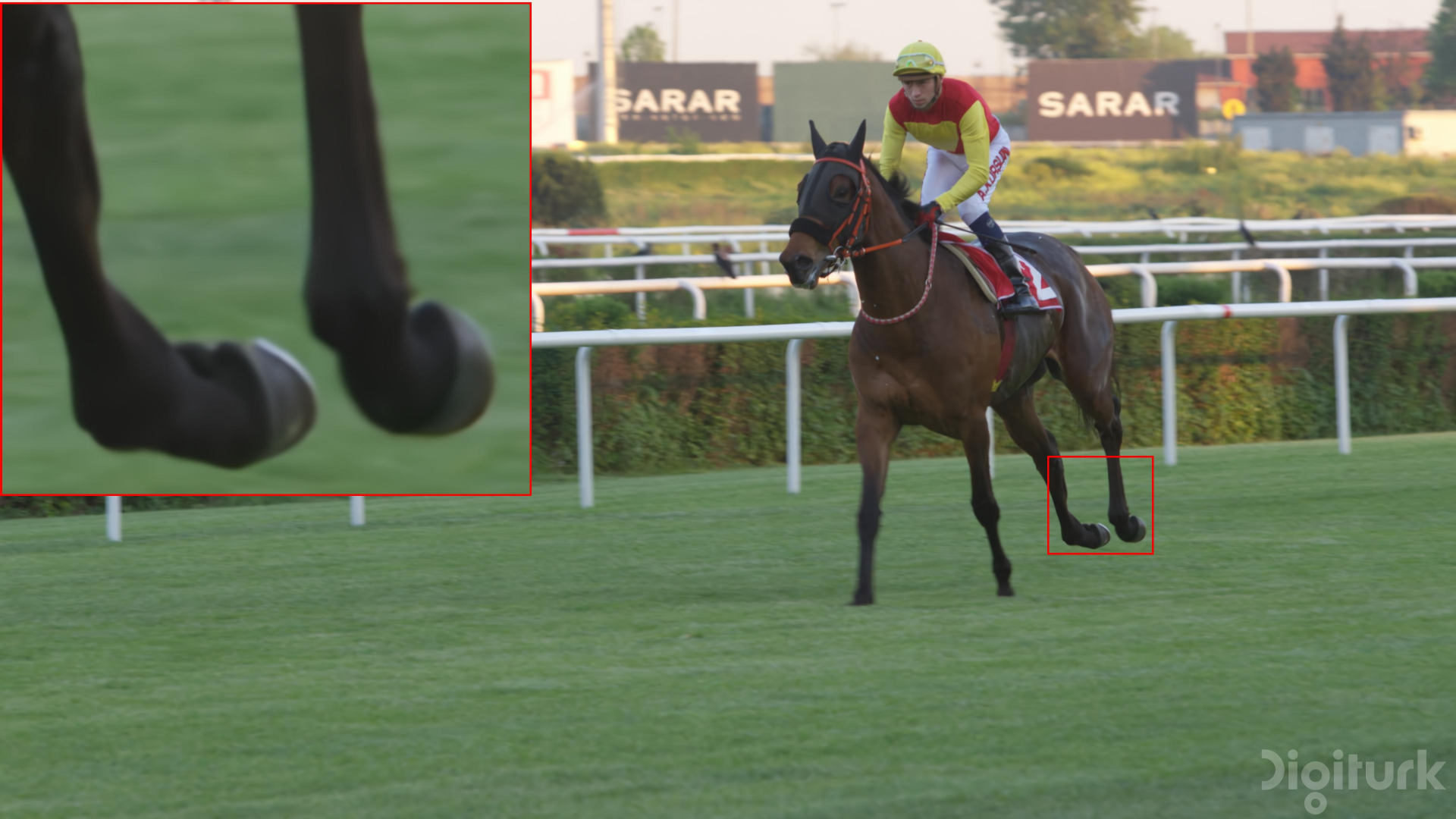}
        \caption{DCVC—RT\\ Bpp:0.034 LPIPS:0.42}
        \label{fig:result2}
    \end{subfigure}

    \vspace{2mm} 

    \begin{subfigure}{0.48\columnwidth}
        \centering
        \includegraphics[width=\linewidth]{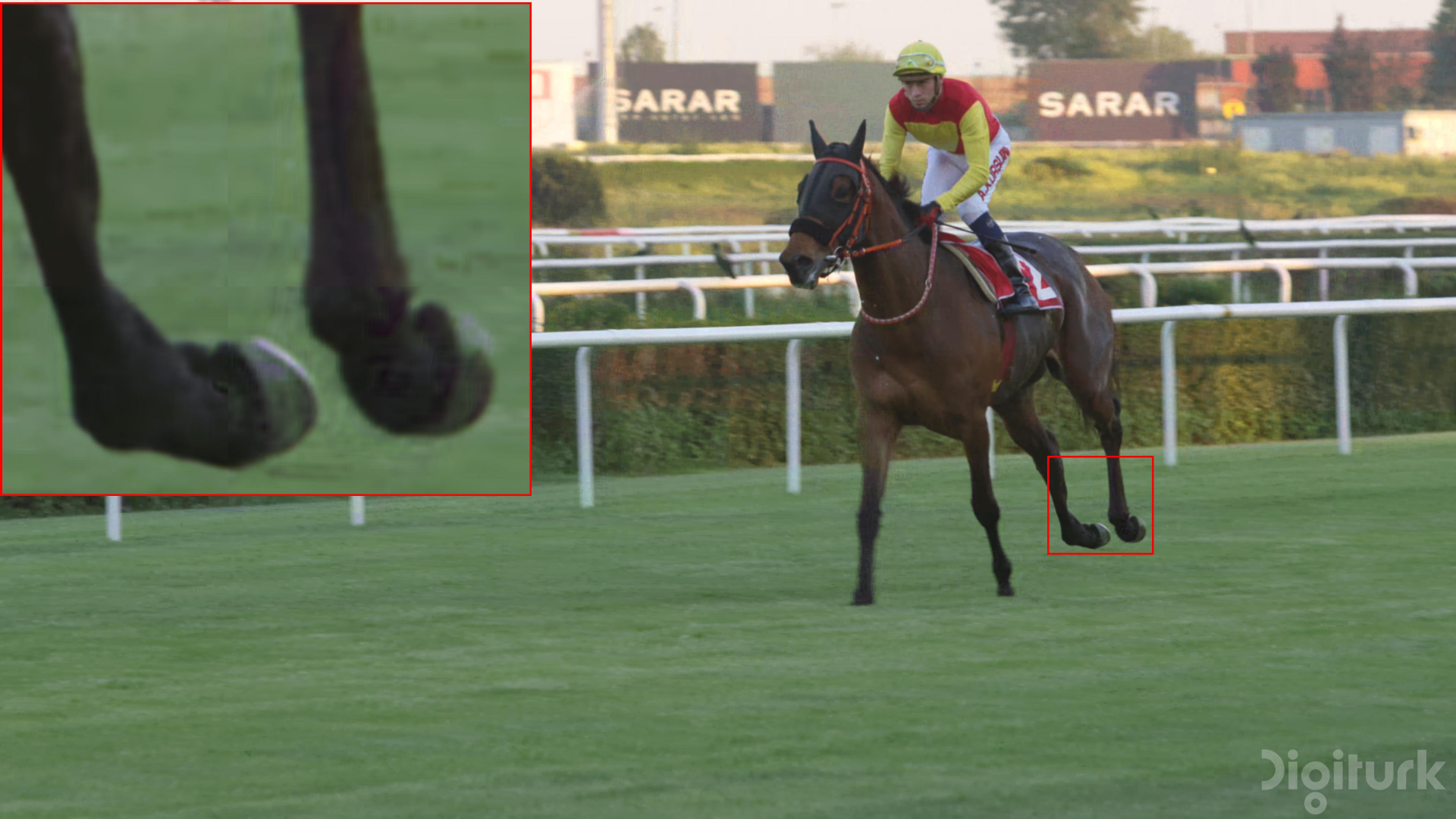}
        \caption{C3\\ Bpp:0.031 LPIPS:0.45}
        \label{fig:result3}
    \end{subfigure}
    \hfill
    \begin{subfigure}{0.48\columnwidth}
        \centering
        \includegraphics[width=\linewidth]{Figures/processed_Ours.pdf}
        \caption{Ours\\ Bpp:0.031 LPIPS:0.29}
        \label{fig:visual}
    \end{subfigure}
    
    \caption{Qualitative comparison on the Jockey sequence.}
    \label{fig:visual_comparison}
\end{figure}

\subsection{Quantitative and Qualitative Results}
\noindent\textbf{Rate-distortion Performance} Fig \ref{fig: RD}. illustrates the RD performance of our proposed method compared to all baselines. It can be observed that in the mid-to-high bitrate range, our method consistently outperforms H.265 across all evaluated metrics and are comparable to state-of-the-art INR methods C3\cite{C3} and HiNeRV\cite{hinerv}. Notably, regarding the LPIPS metric, which is more closely aligned with human visual perception, our approach achieves results comparable to HiNeRV across the entire bitrate spectrum, significantly exceeding other baselines. 

\noindent\textbf{Efficiency Performance} Table \ref{tab:performance_comparison}. presents a comparative analysis of compression efficiency between our proposed method and existing baselines at both low and high bitrates. In terms of decoding FPS and VRAM consumption, our approach demonstrates performance comparable to GSVC\cite{10.1145/3712678.3721876} while substantially outperforming all other baselines. Notably, even when compared to DCVC-RT\cite{DCVC_RTimage}, which is specifically optimized for efficiency, our framework still achieves a remarkable 15$\times$ speedup in decoding FPS and a reduction of 69 \% in VRAM usage.

\noindent\textbf{Visual Results} As illustrated in Fig. \ref{fig:visual_comparison}, our method preserves more fine-grained details compared to DCVC-RT\cite{DCVC_RTimage} and effectively eliminates visual artifacts present in C3\cite{C3}.
\begin{table}[htbp]
\centering
\caption{Ablation Study Results on UVG\cite{UVG} dataset}
\label{tab:ablation_study}
\renewcommand{\arraystretch}{1.3} 
\setlength{\tabcolsep}{4pt}      
\footnotesize                    

\begin{tabular}{@{}lccc@{}}
\toprule
Model & PSNR $\uparrow$ & Dec FPS $\uparrow$ & VRAM (MB) $\downarrow$ \\ \midrule
\textbf{Ours full model}                          & \textbf{36.18} & \textbf{247.21}          & \textbf{912} \\ 
\quad w/o Optical flow                   & 36.02          & 247.19          & 912 \\ 
\quad w/o Multi-stage training           & 35.84          & 247.19          & 912 \\ 
\quad w/o spatial-temporal encoder       & 35.01          & 176.34          & 2087 \\ 
\quad w/o Multi-scale GS                 & 34.29          & 354.16 & 2024 \\ \bottomrule
\end{tabular}
\end{table}

\subsection{Ablation Study}
Table.~\ref{tab:ablation_study} demonstrates the effectiveness of each proposed component in our method. Overall, every module contributes positively to the enhancement of video reconstruction quality. Furthermore, while the inclusion of the multi-scale representation involves a trade-off with decoding speed, our full model still achieves a decoding frame rate remaining far superior to existing state-of-the-art methods. These results validate that our design strikes an optimal balance between visual fidelity and computational efficiency.

%% file: sections/conclusion.tex
\section{Conclusion }
We present a multi-scale 2DGS-based video compression technique that demonstrates superior RD performance over H.265 on 4K resolution videos and order-of-magnitude improvements in decoding FPS and VRAM utilization  compared with other state-of-the-art neural codecs. our work underscores the transformative potential of GS-based representations for future video compression research. Future work could explore the untapped potential of 2DGS-specific compression.